\pdfoutput=1
\documentclass[aps,prl,amsmath,superscriptaddress,amssymb,floatfix,lengthcheck,reprint]{revtex4-1}
\usepackage{graphicx}
\usepackage{textcomp}

\begin{document}
\title{Sideband Cooling Micromechanical Motion to the Quantum Ground State}

\author{J. D. Teufel}
\affiliation{National Institute of Standards and Technology, Boulder, CO 80305, USA}
\author{T. Donner}
\affiliation{JILA, National Institute of Standards and Technology
and the University of Colorado, Boulder, CO 80309, USA}
\author{Dale Li}
\affiliation{National Institute of Standards and Technology, Boulder, CO 80305, USA}
\author{J. W. Harlow}
\affiliation{JILA, National Institute of Standards and Technology
and the University of Colorado, Boulder, CO 80309, USA}
\affiliation{Department of Physics, University of
Colorado, Boulder, Colorado 80309, USA}
\author{M. S. Allman}
\affiliation{National Institute of Standards and Technology, Boulder, CO 80305, USA}
\author{K. Cicak}
\affiliation{National Institute of Standards and Technology, Boulder, CO 80305, USA}
\author{A. J. Sirois}
\affiliation{National Institute of Standards and Technology, Boulder, CO 80305, USA}
\author{J. D. Whittaker}
\affiliation{National Institute of Standards and Technology, Boulder, CO 80305, USA}
\author{K. W. Lehnert}
\affiliation{JILA, National Institute of Standards and Technology
and the University of Colorado, Boulder, CO 80309, USA}
\affiliation{Department of Physics, University of
Colorado, Boulder, Colorado 80309, USA}
\author{R. W. Simmonds}
\affiliation{National Institute of Standards and Technology, Boulder, CO 80305, USA}

\maketitle

\textbf{
The advent of laser cooling techniques revolutionized the study of
many atomic-scale systems. This has fueled progress towards
quantum computers by preparing trapped ions in their motional
ground state \cite{Diedrich1989}, and generating new states of
matter by achieving Bose-Einstein condensation of atomic vapors
\cite{Anderson1995}. Analogous cooling techniques
\cite{Braginsky1992,Kippenberg2008} provide a
general and flexible method for preparing macroscopic objects in
their motional ground state, bringing the powerful technology of
micromechanics into the quantum regime. Cavity opto- or
electro-mechanical systems achieve sideband cooling through the
strong interaction between light and motion
\cite{Braginsky1970,Blair1995,Teufel2008,Thompson2008,Groblacher2009a,Park2009,Lin2009,Schliesser2009,Rocheleau2010,Riviere2010,Li2011}.
However, entering the quantum regime, less than a single quantum
of motion, has been elusive because sideband cooling has not
sufficiently overwhelmed the coupling of mechanical systems to
their hot environments. Here, we demonstrate sideband cooling of
the motion of a micromechanical oscillator to the quantum ground
state. Entering the quantum regime requires a large
electromechanical interaction, which is achieved by embedding a
micromechanical membrane into a superconducting microwave resonant
circuit. In order to verify the cooling of the membrane motion
into the quantum regime, we perform a near quantum-limited
measurement of the microwave field, resolving this motion a factor
of 5.1 from the Heisenberg limit \cite{Braginsky1992}.
Furthermore, our device exhibits strong-coupling allowing coherent
exchange of microwave photons and mechanical
phonons \cite{Teufel2010}. Simultaneously achieving strong
coupling, ground state preparation and efficient measurement sets
the stage for rapid advances in the control and detection of
non-classical states of motion \cite{Bose1997,Mancini1997},
possibly even testing quantum theory itself in the unexplored
region of larger size and mass \cite{Marshall2003}.
The universal ability to connect disparate physical systems
through mechanical motion naturally leads towards future methods
for engineering the coherent transfer of quantum information with
widely different forms of quanta.
}

Mechanical oscillators that are both decoupled from their
environment (high quality factor $Q$) and placed in the quantum
regime could allow us to explore quantum mechanics in entirely new
ways \cite{Bose1997,Mancini1997,Marshall2003,Akram2010,Regal2011}.
For an oscillator to be in the quantum regime, it must be possible
to prepare it in its ground state, to arbitrarily manipulate its
quantum state, and to detect its state near the Heisenberg limit.
In order to prepare an oscillator in its ground state, its
temperature $T$ must be reduced such that
$k_\mathrm{B}T<\hbar\Omega_\mathrm{m}$, where $\Omega_\mathrm{m}$
is the resonance frequency of the oscillator, $k_\mathrm{B}$ is
Boltzmann's constant, and $\hbar$ is the reduced Planck's
constant. While higher resonance frequency modes ($>1$~GHz) can
meet this cooling requirement with conventional refrigeration
($T<50$~mK),  these stiff oscillators are difficult to control and
to detect within their short mechanical lifetimes. One unique
approach using passive cooling has successfully overcome these
difficulties by using a piezoelectric dilatation oscillator
coupled to a superconducting qubit \cite{OConnell2010}.
Unfortunately, this method is incompatible with the broad range of
lower frequency, high Q, flexural mechanical modes. In order to
take advantage of the attractive mechanical properties of these
oscillators, an alternative active cooling method is required, one
that can reduce the oscillator's temperature below that of the
surrounding environment.

Cavity opto- or electro-mechanical systems
\cite{Kippenberg2008} naturally offer a method for
both detecting mechanical motion and cooling a mechanical mode to
its ground state  \cite{Marquardt2007,Wilson2007}.  An object
whose motion alters the resonance frequency $\omega_\mathrm{c}$ of
an electromagnetic cavity experiences a radiation pressure force
governed by the parametric interaction Hamiltonian:
$\hat{H}_\mathrm{int}=\hbar G\hat{n}\hat{x}$, where
$G=d\omega_\mathrm{c}/dx$, $\hat{n}$ is the cavity photon number,
and $\hat{x}$ is the displacement of the mechanical oscillator. By
driving the cavity at a frequency $\omega_\mathrm{d}$, the
oscillator's motion produces upper and lower sidebands at
$\omega_\mathrm{d}\pm \Omega_\mathrm{m}$. Because these sideband
photons are inelastically scattered from the drive field, they
provide a way to exchange energy with the oscillator. If the drive
field is optimally detuned below the cavity resonance $\Delta
\equiv
\omega_{\mathrm{d}}-\omega_{\mathrm{c}}=-\Omega_{\mathrm{m}}$,
photons will be preferentially up-converted to $\omega_\mathrm{c}$
because the photon density of states is maximal there (Fig~1b).  When
an up-converted photon leaves the cavity, it removes the energy of
one mechanical quantum (one phonon) from the motion. Thus, the
mechanical oscillator is damped and cooled via this
radiation-pressure force. Because the mechanical motion is encoded
in scattered photons exiting the cavity, a quantum-limited
measurement of this photon field provides a near
Heisenberg-limited detection of mechanical motion
\cite{Clerk2010}.

While there has been substantial progress in cooling mechanical
oscillators with radiation pressure forces, sideband cooling to the
quantum mechanical ground state has been an outstanding challenge.
Cavity optomechanical systems have realized very large sideband
cooling rates \cite{Thompson2008,Groblacher2009a,Park2009,Lin2009,Schliesser2009,Riviere2010,Li2011}; however, these rates are not
sufficient to overcome the larger thermal heating rates of the mechanical modes.
Because electromechanical experiments use much lower-energy
photons \cite{Braginsky1970,Blair1995,Teufel2008,Rocheleau2010}, they are naturally compatible with
operation below $100$~mK, but have consequently suffered from weak
electromechanical interactions and inefficient detection of the
photon fields.

Here, we present a cavity electromechanical system where a
flexural mode of a thin aluminum membrane is parametrically
coupled to a superconducting microwave resonant circuit. Unlike
previous microwave systems, this device achieves large
electromechanical coupling by concentrating nearly all the
microwave electric fields near the mechanical
oscillator \cite{Teufel2010}. The oscillator is a $100$~nm thick
aluminum membrane with a diameter of 15~\textmu m, suspended $50$~nm
above a second aluminum layer on a sapphire substrate \cite{Cicak2010} (see Fig.
1). These two metal layers form a variable parallel-plate
capacitor that is shunted by a $12$~nH spiral inductor. This
combination of capacitor and inductor creates a microwave cavity
whose resonance frequency depends on the mechanical displacement
of the membrane and is centered at $\omega_{\mathrm{c}}=2\pi
\times 7.54$~GHz. The device is operated in a dilution
refrigerator at $15$~mK, where aluminum is superconducting, and
the microwave cavity has a total energy decay rate of $\kappa \approx
2 \pi \times 200$~kHz.  As expected from the dimensions
of the membrane, $\Omega_{\mathrm{m}}=2 \pi \times 10.56$~ MHz,
and we find an intrinsic damping rate of $\Gamma_{\mathrm{m}}=2
\pi \times 32$~Hz, resulting in a mechanical quality factor $Q_\mathrm{m}=\Omega_\mathrm{m}/\Gamma_\mathrm{m}=3.3 \times 10^5$. The oscillator mass $m=48$~pg implies that the zero-point motion is $x_\mathrm{zp}=\sqrt{\hbar/(2m\Omega_\mathrm{m})}=4.1$~fm.  With a ratio of
$\Omega_{\mathrm{m}}/\kappa>50$, our system is deep in the
resolved-sideband regime and perfectly suited for sideband cooling to the mechanical ground state \cite{Marquardt2007,Wilson2007}.

To measure the mechanical displacement, we apply a microwave field, which is detuned below the cavity resonance frequency by $\Delta=-\Omega_\mathrm{m}$, through heavily attenuated
coaxial lines to the feed line of our device. The
upper sideband at $\omega_\mathrm{c}$ is amplified with a custom-built Josephson
parametric amplifier (JPA) \cite{Castellanos-Beltran2008,Teufel2009} followed by a
low-noise cryogenic amplifier, demodulated at room temperature,
and finally monitored with a spectrum analyzer. The thermal motion of the membrane creates an easily resolvable peak in the microwave noise spectrum. As described previously\cite{Teufel2009}, this
measurement scheme constitutes a nearly shot-noise-limited
microwave interferometer with which we can measure mechanical
displacement with minimum added noise close to fundamental
limits.

In order to calibrate the demodulated signal to the membrane's
motion, we measure the thermal noise spectrum while varying the
cryostat temperature (Fig.~1c).   Here a weak microwave drive
($\sim 3$ photons in the cavity) is used in order to ensure that
radiation pressure damping and cooling effects are negligible.
When $\Omega_\mathrm{m}\gg \kappa \gg \Gamma_\mathrm{m}$ and
$\Delta=-\Omega_\mathrm{m}$, the displacement spectral density
$S_x$ is related to the observed microwave noise spectral density
$S$ by:  $S_x=2 (\kappa\Omega_\mathrm{m}/G\kappa_\mathrm{ex})^2
S/P_\mathrm{o}$, where $\kappa_\mathrm{ex}$ is the coupling rate
between the cavity and the feed line, and $P_\mathrm{o}$ is the
power of the microwave drive at the output of the cavity.
According to equipartition, the area under the resonance curve of
displacement spectral density $S_x$ must be proportional to the
effective temperature of the mechanical mode. This calibration
procedure allows us to convert the sideband in the microwave power
spectral density to a displacement spectral density and to extract
the thermal occupation of the mechanical mode. In Fig.~1c we show
the number of thermal quanta in the mechanical resonator as a
function of $T$. The linear dependence of the integrated power
spectral density with temperature shows that the mechanical mode
equilibrates with the cryostat even for the lowest achievable
temperature of $15$~mK. This temperature corresponds to a thermal
occupancy $n_\mathrm{m}=30$, where $n_\mathrm{m}=[\exp(\hbar
\Omega_{\mathrm{m}}/k_{\mathrm{B}}T)-1]^{-1}$. The calibration
determines the electromechanical coupling strength $G/2\pi=49 \pm
2$~MHz/nm. With the device parameters, we can investigate both the
fundamental sensitivity of our measurement as well as the effects
of radiation pressure cooling.

The total measured displacement noise results from two sources:
the membrane's actual mean-square motion $S_x^\mathrm{th}$ and the
\emph{apparent} motion $S_x^\mathrm{imp}$ due to imprecision of
the measurement. Fig.~2a demonstrates how the use of low-noise
parametric amplification significantly lowers $S_x^\mathrm{imp}$,
resulting in a reduction in the white-noise background by a factor
of more than $30$.  This greatly increases the signal-to-noise
ratio of the membrane's thermal motion, reducing the required
integration time to resolve the thermal peak by a factor of
$1000$. To investigate the measurement sensitivity in the presence
of dynamical backaction, we regulate the cryostat temperature at
$20$~mK and increase the amplitude of the detuned microwave drive
while observing modifications in the displacement spectral
density.   We quantify the strength of the drive by the resulting
number of photons $n_\mathrm{d}$ in the microwave cavity.  As
shown in Fig.~2b, the measurement imprecision
$S_\mathrm{x}^\mathrm{imp}$ is inversely proportional to
$n_\mathrm{d}$.  At the highest drive power ($n_\mathrm{d} \approx
10^5$),  the absolute displacement sensitivity is $5.5 \times
10^{-34}$~m$^2$/Hz.

As expected, the increased drive power also damps and cools the
mechanical oscillator \cite{Braginsky1992,
Marquardt2007,Wilson2007}.  The total mechanical dissipation rate
$\Gamma_\mathrm{m}'$ is the sum of the intrinsic dissipation
$\Gamma_\mathrm{m}$ and the radiation-pressure-induced damping
resulting from scattering photons to the upper/lower sideband
$\Gamma=\Gamma_\mathrm{+}-\Gamma_\mathrm{-}$, where
$\Gamma_\mathrm{\pm}=4 g^2 \kappa/[\kappa^2+4(\Delta \pm
\Omega_\mathrm{m})^2]$.  Here, $g$ is the coupling rate between
the cavity and the mechanical mode, which depends on the amplitude
of the drive: $g=G x_\mathrm{zp}\sqrt{n_\mathrm{d}}$.  Fig.~2c
shows the measured values of $\kappa$, $g$ and
$\Gamma_\mathrm{m}'$ as the drive increases.  The
radiation-pressure damping of the mechanical oscillator becomes
pronounced above a cavity drive amplitude of approximately 75
photons, at which point $\Gamma=\Gamma_\mathrm{m}$ and the
mechanical linewidth has doubled.

While the absolute value of the displacement imprecision decreases
with increasing power, the visibility of the thermal mechanical
peak no longer improves once the radiation-pressure force becomes
the dominant dissipation mechanism for the membrane. By expressing
the imprecision as equivalent thermal quanta of the oscillator
$n_\mathrm{imp}=\Gamma_\mathrm{m}'S_x^\mathrm{imp}/8
x_\mathrm{zp}^2$, we see that the visibility of the thermal noise
above the imprecision no longer improves once the drive is much
greater than $n_\mathrm{d}\approx 100$ (Fig. 2d). This is because
a linear decrease in $ S_x^\mathrm{imp}$ is balanced by a linear
increase in $\Gamma_\mathrm{m}'$ due to radiation-pressure
damping.  The asymptotic value of $n_\mathrm{imp}$ is a direct
measure of the efficiency of the microwave measurement. Ideally,
for a lossless circuit, a quantum-limited microwave measurement
would imply $n_\mathrm{imp}=1/4$.  The incorporation of the
low-noise JPA improves $n_\mathrm{imp}$ close to this ideal limit,
reducing the asymptotic value of  $n_\mathrm{imp}$ from $70$ to
$1.9$ quanta. This level of sensitivity is crucial, as we will now
use this measurement to resolve the residual thermal motion of the
membrane as it is cooled into the quantum regime.

Beginning from a cryostat temperature of $20$~mK and a thermal occupation of
$n_\mathrm{m}^\mathrm{T}=40$ quanta, the fundamental mechanical
mode of the membrane is cooled by the radiation-pressure forces.
Figure~3a shows the displacement spectral density of the motional
sideband as $n_\mathrm{d}$ is increased from 18 to 4,500 photons
along with fits to a Lorentzian lineshape (shaded area).  As described above,
this increased drive results in three effects on the spectra:
lower noise floor, wider resonances and smaller area.  As it is
the area that corresponds to the mean-square motion of the
membrane, it directly measures the effective temperature of the
mode.   At a drive intensity that corresponds to 4,000 photons in
the cavity, the thermal occupation is reduced below one quantum of
mechanical motion, entering the quantum regime.

Observing the noise spectrum over a broader frequency range
reveals that there is also a second Lorentzian peak with linewidth
$\kappa$ whose area corresponds to the finite thermal occupation
$n_\mathrm{c}$ of the cavity.  Over a broad frequency range it is
no longer valid to evaluate the cavity parameters at a single
frequency to infer the spectrum in units of $S_x$.  Instead,
Fig.~3b shows the noise spectrum in units of sideband power
normalized by the power at the drive frequency, $S/P_\mathrm{o}$.
These two sources of noise originating from either the mechanical
or the electrical mode interfere with each other and result in
noise squashing \cite{Rocheleau2010} and eventually normal-mode
splitting \cite{Dobrindt2008} once $2g>\kappa/\sqrt{2}$.  Using a
quantum-mechanical description applied to our circuit
\cite{Rocheleau2010,Clerk2010}, the expected noise spectrum is
\small
\begin{equation}
S/\hbar \omega=\frac{1}{2}+n_\mathrm{add}+\frac{2\kappa_\mathrm{ex}\left[ \kappa n_\mathrm{c} (\Gamma_\mathrm{m}^2+ 4\delta^2)+4 \Gamma_\mathrm{m} n_\mathrm{m}^\mathrm{T} g^2\right] }{\left|4 g^2+\left(\kappa+2j (\delta+\widetilde{\Delta})\right)\left(\Gamma_\mathrm{m}+2j \delta\right)\right|^2}\
\end{equation}
\normalsize where $\delta= \omega-\Omega_\mathrm{m}$,
$\widetilde{\Delta}=\omega_\mathrm{d}+\Omega_\mathrm{m}-\omega_\mathrm{c}$,
and $n_\mathrm{add}$ is added noise of the microwave measurement
expressed as an equivalent number of microwave photons.  Fig.~3b
shows the measured spectra and corresponding fits (shaded region)
to Eq.~1 as the electromechanical system evolves first into the
quantum regime ($n_\mathrm{m}, n_\mathrm{c}<1$) and then into the
strong-coupling regime ($2g>\kappa/2$).  The results are
summarized in Fig.~3c, where the thermal occupancy of both the
mechanical and electrical modes are shown as a function of
$n_\mathrm{d}$.  For low drive power, the cavity shows no
resolvable thermal population (to within our measurement
uncertainty of 0.05 quanta) as expected for a $7.5$~GHz mode at
$20$~mK.  While it is unclear whether the observed population at higher
drive power is a consequence of direct heating of the substrate,
heating of the microwave attenuators preceding the circuit, or
intrinsic cavity frequency noise, we have determined that it is
not the result of frequency or amplitude noise of our microwave
generator, as this noise is reduced far below the microwave shot-noise level with narrow-band filtering and cryogenic attenuation
(see Supplementary Information). Sideband cooling can never reduce
the occupancy of the mechanical mode below that of the cavity.
Therefore, in order for the system to access the quantum regime,
the thermal population of the cavity must remain less than one
quantum. Assuming $\Omega_\mathrm{m}\gg \kappa$, the final
occupancy of a mechanical mode is \cite{Dobrindt2008}
\begin{equation}
n_\mathrm{m}=n_\mathrm{m}^\mathrm{T}\left( \frac{\Gamma_\mathrm{m}}{\kappa}\frac{4g^2+\kappa^2}{4g^2+\kappa \Gamma_\mathrm{m}}\right)+n_\mathrm{c}\left(\frac{4g^2}{4g^2+\kappa \Gamma_\mathrm{m}}\right).
\end{equation}
This equation shows that for moderate coupling ($\sqrt{\kappa
\Gamma_\mathrm{m}}\ll g \ll \kappa$) the cooling of the mechanical
mode is linear in the number of drive photons.  Beyond this
regime, the onset of normal-mode splitting abates  further
cooling. Here the mechanical cooling rate becomes limited not by
the coupling between the mechanical mode and the cavity, but
instead by the coupling rate $\kappa$ between the cavity and its
environment \cite{Dobrindt2008}.   Thus, the final occupancy of
the mechanical mode can never be reduced to lower than
$n_\mathrm{m}^\mathrm{T}\Gamma_\mathrm{m}/\kappa$, and a stronger
parametric drive will only increase the rate at which the thermal
excitations Rabi oscillate between the cavity and mechanical
modes.   For our device we achieve the desired hierarchy:  as the
coupling is increased, we first cool to the ground state and then
enter the strong-coupling regime
($n_\mathrm{m}^\mathrm{T}\Gamma_\mathrm{m}<\kappa<g$).  Once
$n_\mathrm{d}$ exceeds $2 \times 10^4$, the mechanical occupancy
converges toward the cavity population, reaching a minimum of
$0.34 \pm 0.05$ quanta. At the highest power drive power
($n_\mathrm{d}=2 \times 10^5$) the mechanical mode has hybridized
with the cavity, resulting in the normal-mode splitting
characteristic of the strong-coupling regime \cite{Teufel2010}.
This level of coupling is required to utilize the hybrid system
for quantum information processing, as it is only in the
strong-coupling regime that a quantum state may be manipulated
faster than it decoheres from the coupling of either the
electromagnetic or mechanical modes to the environment.

Together the measurements shown in Fig.~2 and 3 quantify the
overall measurement efficiency of the system.  The Heisenberg
limit requires that a continuous displacement measurement is
necessarily accompanied by a backaction force
\cite{Braginsky1992,Clerk2010, Schliesser2009}, such that
$\sqrt{S_x^\mathrm{imp} S_F}\geq \hbar$, where $S_F$ is the force
noise spectral density. From the thermal occupancy and damping
rate of the mechanical mode, we extract the total force spectral
density $S_F=4\hbar
\Omega_\mathrm{m}m\Gamma_\mathrm{m}'(n_\mathrm{m}+1/2)$.  This
places a conservative upper bound on the quantum backaction by
assuming that it alone is responsible for the finite occupancy of
the mechanical mode.  This experiment achieves the closest
approach to Heisenberg-limited displacement detection to date
\cite{Clerk2010,Riviere2010} with a lowest imprecision-backaction
product $\sqrt{S_\mathrm{x}^\mathrm{imp} S_\mathrm{F}}=
4\hbar\sqrt{n_\mathrm{imp}(n_\mathrm{m}+1/2)}=(5.1 \pm 0.4
)\hbar$.  Thus, this mechanical device simultaneously demonstrates
ground-state preparation, strong-coupling and near quantum-limited
detection.

Looking forward, this technology offers a feasible route to achieve many of the longstanding goals for quantum \emph{mechanical} systems.  These prospects include a direct measurement of the zero-point motion, observation of the fundamental asymmetry between the rate of emission and absorption of phonons \cite{Diedrich1989}, quantum nondemolition measurements \cite{Braginsky1992} and generation of entangled states of mechanical motion \cite{Bose1997,Mancini1997}.  Furthermore, combining this device with a single-photon source and detector (such as a superconducting qubit \cite{Hofheinz2009,OConnell2010}) would enable preparation of arbitrary quantum states of mechanical motion as well as observation of a single excitation as it Rabi oscillates between a $7$~GHz photon and a $10$~MHz phonon \cite{Akram2010}.  Because the interaction between the mechanical mode and the cavity is parametric, the coupling strength is inherently tunable and can be turned on and off quickly.  Thus, once a quantum state is transfered into the mechanical mode, it can be stored there for a time $\tau_\mathrm{th}=1/(n_\mathrm{m}^\mathrm{T}\Gamma_\mathrm{m})>100$~\textmu s before absorbing one thermal phonon from its environment. As this timescale is much longer than typical coherence times of superconducting qubits, mechanical modes offer the potential for delay and storage of quantum information.  Lastly, because mechanical oscillators can couple to light of any frequency, they could serve as a unique intermediary that transfers quantum information between the microwave and optical domains \cite{Regal2011}.

These measurements demonstrate the power of sideband techniques to cool a macroscopic ($\sim 10^{12}$ atoms) mechanical mode, beyond what is feasible with conventional refrigeration techniques, into the quantum regime.  These broadly applicable methods for state preparation, manipulation and detection, pave the way to access the quantum nature of  a wide class of long-lived mechanical oscillators. Through the strong interaction between photons and phonons, mechanical systems can now inherit the experimental and theoretical power of quantum optics, opening the field of quantum acoustics.

\section{Acknowledgements} We thank A. W. Sanders for taking the
micrograph in Fig. ~1a and thank the JILA instrument shop for fabrication and design of the cavity filter.  This work was financially supported by NIST and the DARPA QuASAR program.  T.D. acknowledges support from the Deutsche Forschungsgemeinschft (DFG).
Contribution of the U.S. government, not subject to copyright.

\section{Author Information}
Reprints and permissions information is
available at www.nature.com/reprints. The authors declare no
competing financial interests. Correspondence and requests for
materials should be addressed to J.D.T (john.teufel@nist.gov).

\pagebreak

\begin{figure*}[h] 
\includegraphics[width=183mm]{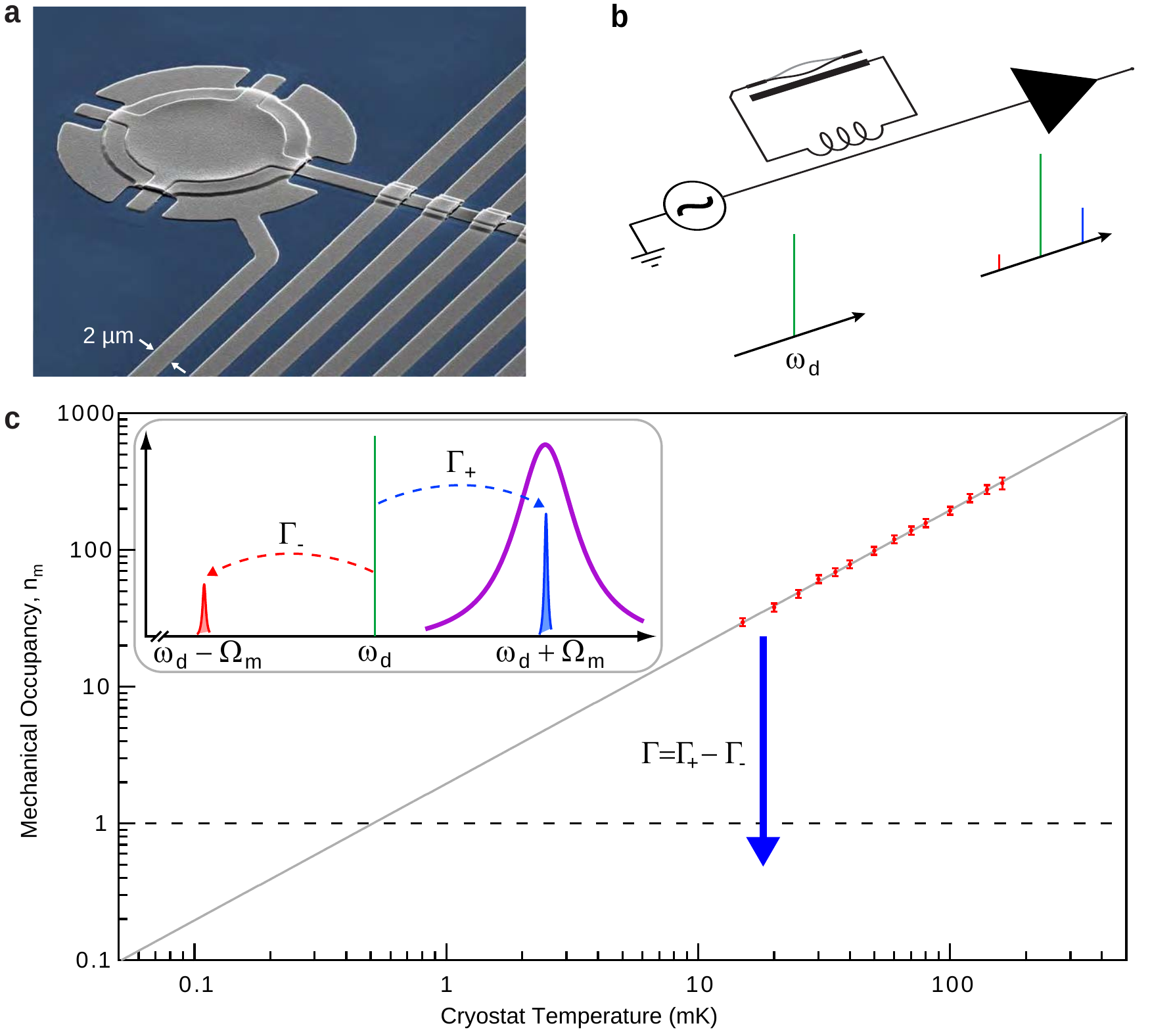}
\caption{\textbf{Schematic description of the experiment.}
\textbf{a},  Colorized scanning electron micrograph
showing the aluminum (grey) electromechanical circuit fabricated on a sapphire (blue) substrate, in which a 15~\textmu m diameter membrane is lithographically suspended 50~nm above a lower electrode. The membrane's motion modulates the capacitance, and hence, the resonance frequency of the superconducting microwave circuit. \textbf{b}, A  coherent microwave drive inductively coupled to the circuit acquires modulation sidebands due to the thermal motion of the membrane.  The upper sideband is amplified with a nearly quantum-limited Josephson parametric amplifier within the cryostat.   \textbf{c}, The microwave power in the upper sideband provides a direct measurement of the thermal occupancy of the mechanical mode, which may be calibrated \emph{in situ} by varying the temperature of the cryostat.  The mechanical mode shows thermalization with the cryostat at all temperatures, yielding a minimum thermal occupancy of $30$ mechanical quanta without employing sideband-cooling techniques. The inset illustrates the concept of sideband cooling.  When the circuit is excited with a detuned microwave drive such that $\Delta=-\Omega_\mathrm{m}$, the narrow line shape of the electrical resonance ensures that photons are preferentially scattered to higher energy, providing a cooling mechanism for the membrane.}
\end{figure*}
\pagebreak
\begin{figure*}[h] 
\includegraphics{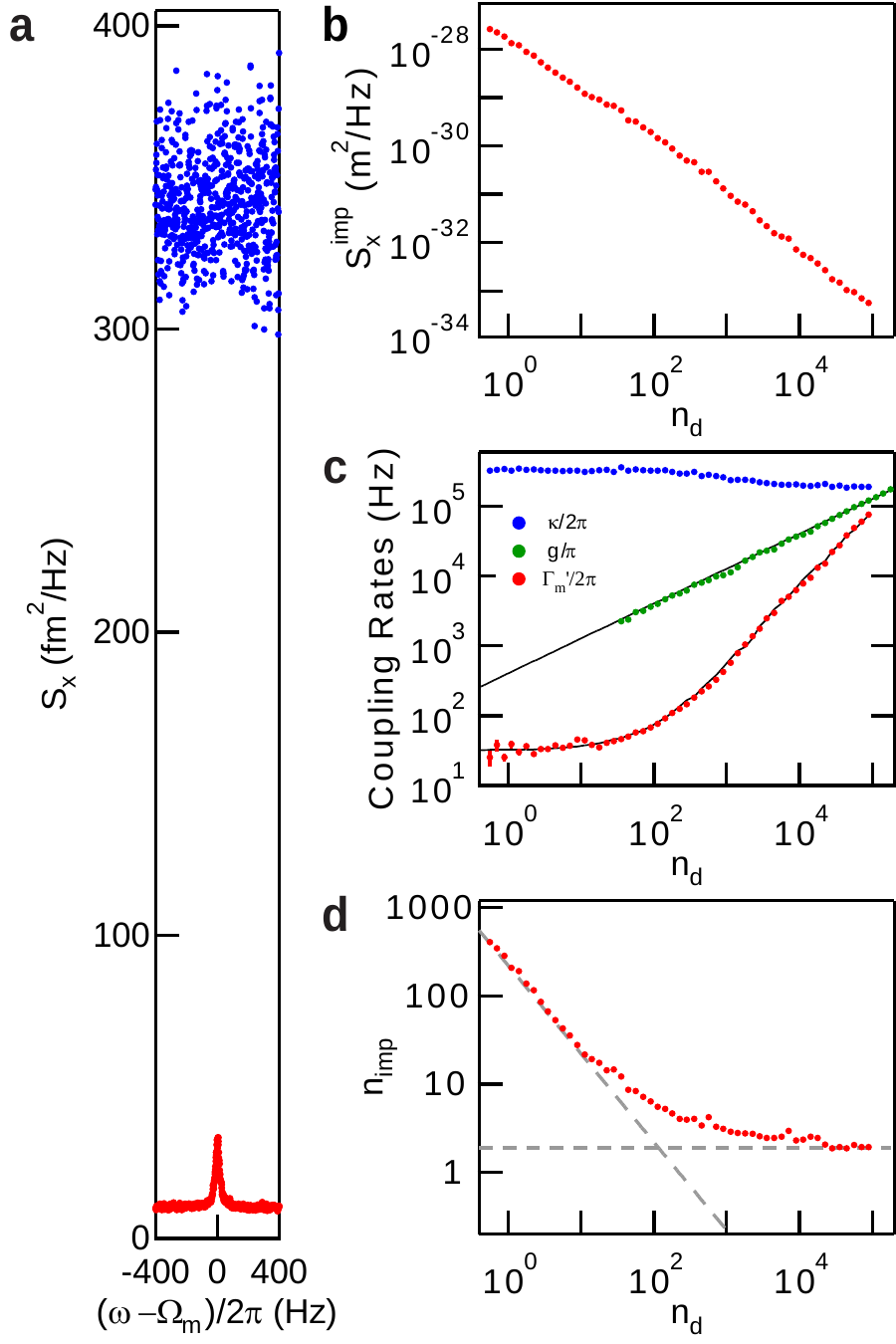}
\caption{  \textbf{Displacement sensitivity in the presence of radiation-pressure damping.} \textbf{a}, The displacement spectral density measured with (red) and without (blue) the Josephson parametric amplifier.  As the parametric amplifier greatly reduces the total noise of the microwave measurement, the time required to resolve the thermal motion is reduced by a factor of  $1000$.  \textbf{b}, As the microwave drive power is increased, the absolute displacement sensitivity, $S_\mathrm{x}^\mathrm{imp}$ improves, reaching a minimum of $5.5 \times 10^{-34}$~m$^2$/Hz at the highest power.  \textbf{c}, The parametric coupling $g$ between the microwave cavity and the mechanical mode increases as $\sqrt{n_{\mathrm{d}}}$.  This coupling damps the mechanical mode from its intrinsic linewidth of $\Gamma_{\mathrm{m}}=2 \pi \times 32$~Hz until it is increased to that  of the microwave cavity $\kappa$. \textbf{d},  The relative measurement imprecision, in units of mechanical quanta, depends on the product of $S_\mathrm{x}^\mathrm{imp}$ and $\Gamma_\mathrm{m}'$.  Thus, once the power is large enough that radiation-pressure damping overwhelms the intrinsic mechanical dissipation, $n_\mathrm{imp}$  asymptotically approaches a constant value ($n_\mathrm{imp}=1.9$), which is a direct measure of the overall efficiency of the photon measurement.}
\end{figure*}
\pagebreak
\begin{figure*}[h] 
\includegraphics[width=183mm]{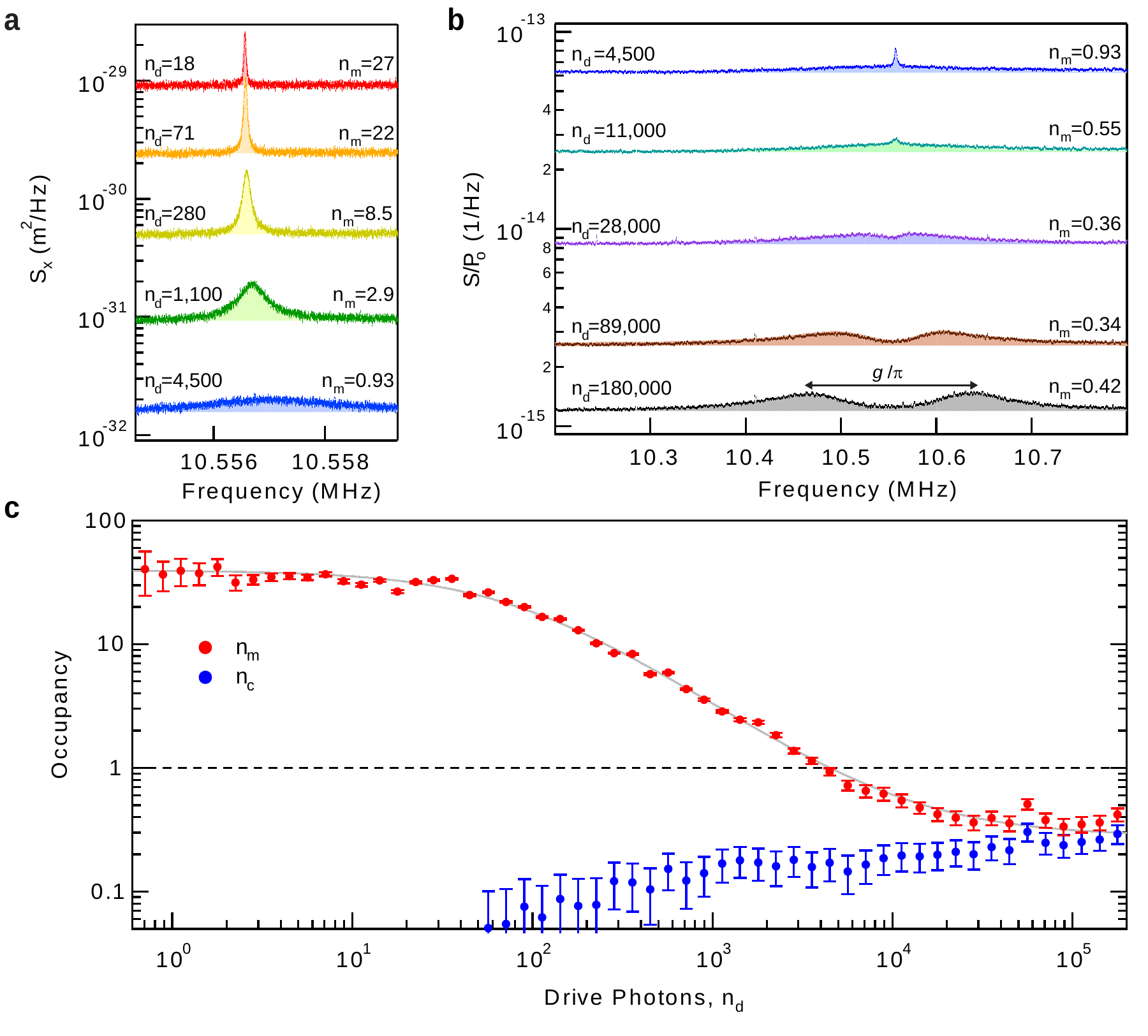}
\caption{ \textbf{Sideband cooling the mechanical mode to the ground state.}
\textbf{a}, The displacement noise spectra and Lorentzian fits (shaded region) for five different drive powers.  With higher power, the mechanical mode is both damped (larger linewidth) and cooled (smaller area) by the radiation pressure forces. \textbf{b}, Over a broader frequency span, the normalized sideband noise spectra clearly show both the narrow mechanical peak and a broader cavity peak due to finite occupancy of the mechanical and electrical modes, respectively.  A small, but resolvable, thermal population of the cavity appears as the drive power increases, setting the limit for the final occupancy of the coupled optomechanical system. At the highest drive power, the coupling rate between the mechanical oscillator and the microwave cavity exceeds the intrinsic dissipation of either mode, and the system hybridizes into optomechanical normal modes. \textbf{c}, Starting in thermal equilibrium with the cryostat at $T=20$~mK, sideband cooling reduces the thermal occupancy of the mechanical mode from $n_{\mathrm{m}}=40$ into the quantum regime, reaching a minimum of $n_{\mathrm{m}}=0.34 \pm 0.05$. These data demonstrate that the parametric interaction between photons and phonons can initialize the strongly coupled, electromechanical system in its quantum ground state. }
\end{figure*}
\pagebreak

\newpage
\newpage
\clearpage
\renewcommand{\thefigure}{\textbf{S}\arabic{figure}}
\renewcommand{\theequation}{$\mathrm{S} $\arabic{equation}}
\setcounter {figure} {0}
\setcounter {equation} {0}
\begin{center}
\begin{widetext}
\large{\textbf{Supplementary Information for ``Sideband Cooling Micromechanical Motion to the Quantum Ground State''}}
\vspace{.2in}
\end{widetext}
\end{center}

\section{Noise spectrum of an optomechanical system}

A mechanical degree of freedom that parametrically couples to the cavity resonance frequency modifies the power emerging from the cavity by scattering photons to the upper or lower mechanical sidebands.  To calculate the full noise spectrum of the optomechanical system, we follow the general method of input-output theory \cite{S_Walls1994}.  We define $g=G x_\mathrm{zp}\sqrt{n_\mathrm{d}}$, where $G=d\omega_\mathrm{c}/dx$, $x_\mathrm{zp}=\sqrt{\hbar/2 m \Omega_\mathrm{m}}$, $m$ is the mass, $\omega_\mathrm{c}$ is the cavity resonance frequency, $\Omega_\mathrm{m}$ is the mechanical resonance frequency and $n_\mathrm{d}$ is the number of photons in the cavity due to a drive at frequency $\omega_\mathrm{d}$.  Furthermore, we define the response functions of the mechanical and cavity modes as $\chi_\mathrm{c}^{-1}=\kappa/2+j (\delta+\widetilde{\Delta})$ and  $\chi_\mathrm{m}^{-1}=\Gamma_\mathrm{m}/2+j \delta$, where $\Gamma_\mathrm{m}$ is the mechanical dissipation rate, $\kappa$  is the cavity dissipation rate, $\delta=\omega-\Omega_\mathrm{m}$, $\widetilde{\Delta}=\omega_\mathrm{d}-\omega_\mathrm{c}+\Omega_\mathrm{m}$ and $j=\sqrt{-1}$. $\kappa$ is total cavity dissipation rate due to both the intentional coupling to the transmission line $\kappa_\mathrm{ex}$ and the intrinsic losses $\kappa_0$.   From these parameters, we define the optomechanical self-energy \cite{S_Marquardt2007,S_Clerk2010} as a function of $\delta$:
\begin{align}
 \Sigma(\delta) &=-j g^2 \left[\chi_\mathrm{c}(\delta)-\chi_\mathrm{c}^*(\delta+2\Omega_\mathrm{m})\right] \\
& \approx -j g^2 \chi_\mathrm{c}(\delta)
\end{align}
 The approximation assumes that the drive is near the optimal detuning for cooling ($|\widetilde{\Delta}| \ll\Omega_\mathrm{m}$) and the system is sufficiently in the good-cavity limit ($\Omega_\mathrm{m} \gg \kappa$) such that the cavity response at $(\delta+2\Omega_\mathrm{m})$ may be neglected.  Now the effective mechanical response function $\widetilde{\chi}_\mathrm{m}$  including the optomechanical effects is:
\begin{align}
\widetilde{\chi}_\mathrm{m} &=\frac{\chi_\mathrm{m}}{1+j \chi_\mathrm{m}\Sigma} \\
& \approx \frac{\chi_\mathrm{c}^{-1}}{g^2+\chi_\mathrm{m}^{-1}\chi_\mathrm{c}^{-1}}
\end{align}

The noise at the output of the cavity is characterized by the noise operator $\hat{b}_\mathrm{out}$, which is related to the cavity field operator $\hat{a}$ by  $\hat{b}_\mathrm{out}=\sqrt{\beta \kappa_\mathrm{ex}}\hat{a}$. $\beta$ is a dimensionless factor that depends on the geometry.  Our circuit (shown schematically in Fig.~S1) couples power from the cavity equally to the output and back to the input so here $\beta=1/2$.  In principle, this fraction could be engineered by coupling asymmetrically to the input and the output, or by using a single port cavity ($\beta=1$).
\begin{figure}[!th]
\includegraphics[width=\columnwidth]{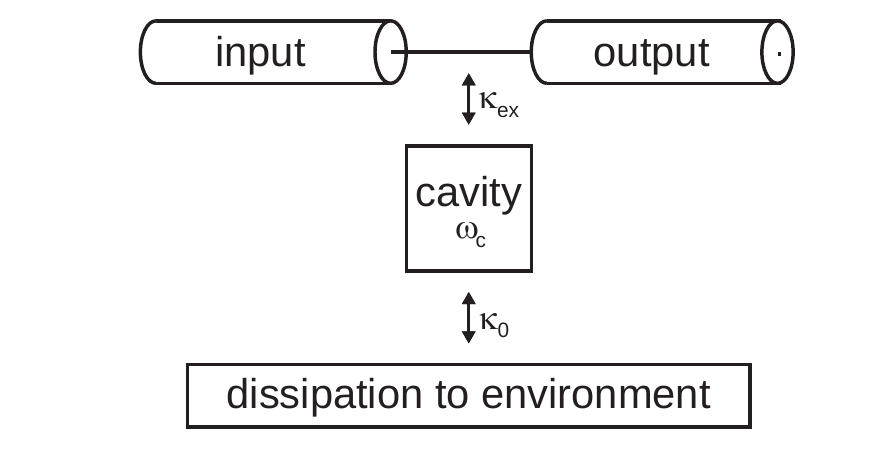}
\caption{\label{supfig1} \textbf{Cavity coupling block diagram.} }
\end{figure} 
\begin{figure*}[!bt]
\includegraphics[width=\textwidth]{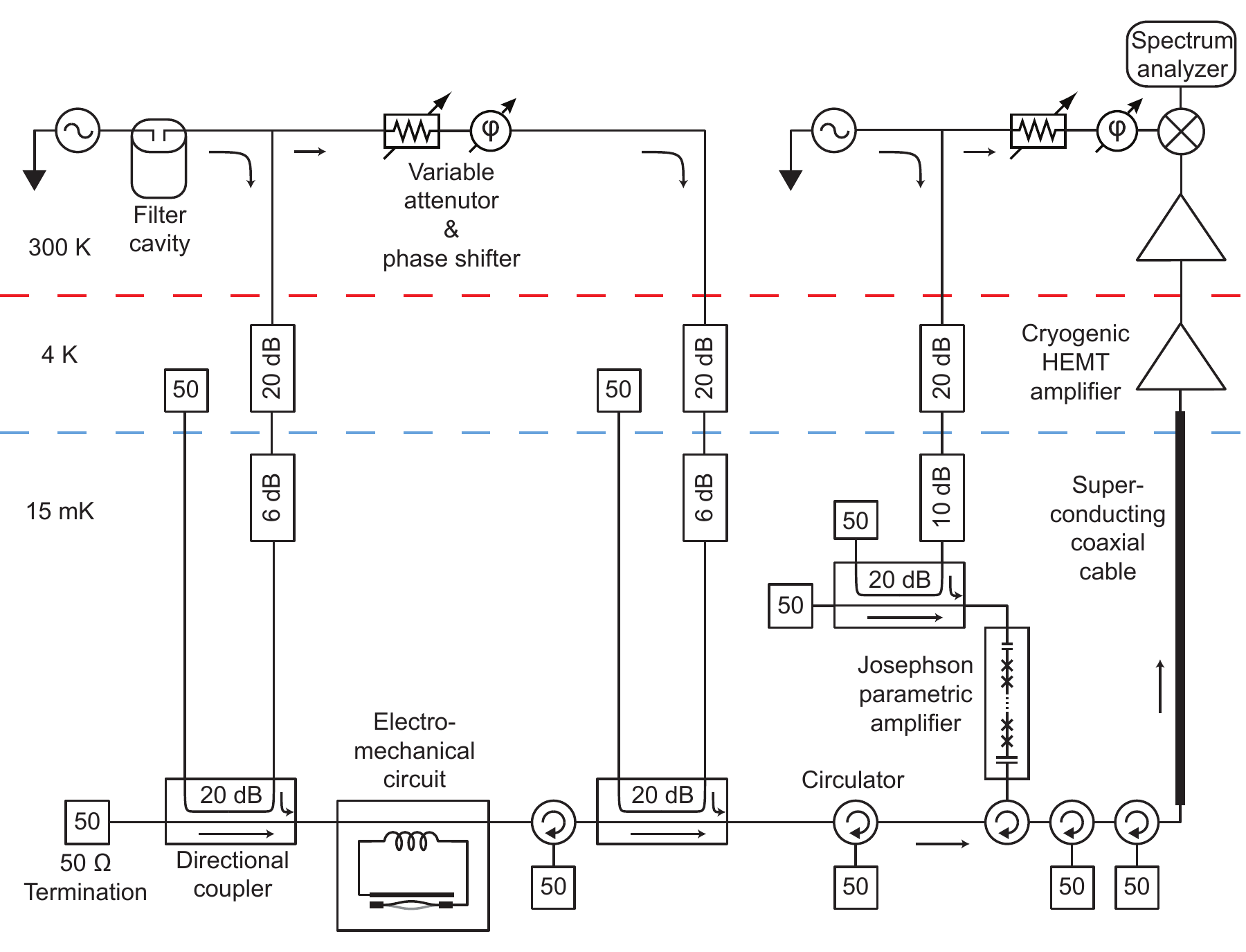}
\caption{\label{supfig2} \textbf{Detailed schematic diagram.} A microwave generator creates a tone at the drive frequency.  This signal is filtered with a resonant cavity at room temperature and split into two arms. The first arm excites the cavity through approximately 53~dB of cryogenic attenuation. In order to avoid saturating the low-noise amplifier with the microwave drive tone, the second arm  is used to cancel the drive before amplification.  A computer-controlled variable attenuator and phase shifter are run in a feedback loop to maintain cancellation at the part per million level.  A second microwave generator is used to provide the pump tone for the Josephson parametric amplifier (JPA) as well as the reference oscillator for the mixer.  This pump tone is $1.3$~MHz above $\omega_\mathrm{c}$ so that the JPA is operated as a non-degenerate parametric amplifier, which measures both quadratures of the electromagnetic filed at the upper sideband frequency. The last stage of attenuation on all lines occurs inside a $20$~dB directional coupler, which allows us to minimize the microwave power dissipated on the cold stage of the cryostat. The JPA is a reflection amplifier; a signal incident on the strongly coupled port of the JPA is reflected and amplified. A cryogenic circulator is used to separate the incident and reflected waves, defining the input and output ports of the JPA. The other circulators are used to isolate the cavity from the noise emitted from the amplifier's input.} 
\end{figure*}
Following directly the theoretical analysis of previous work \cite{S_Rocheleau2010,S_Clerk2010}, we consider the noise operators $\hat{\eta}_\mathrm{m}$ and $\hat{\eta}_\mathrm{c}$ associated with the mechanical and cavity modes respectively, which satisfy the relations $\langle\hat{\eta}_\mathrm{m}^\dagger\hat{\eta}_\mathrm{m}\rangle=n_\mathrm{m}^T$ and $\langle\hat{\eta}_\mathrm{c}^\dagger\hat{\eta}_\mathrm{c}\rangle=n_\mathrm{c}$.  Thus, the output noise is \cite{S_Rocheleau2010}
\begin{align*}
\hat{b}_\mathrm{out}= & -\sqrt{\beta \kappa_\mathrm{ex}}\chi_\mathrm{c} \sqrt{\kappa}\left(1-g^2\widetilde{\chi}_\mathrm{m}\chi_\mathrm{c}\right) \hat{\eta}_\mathrm{c}  \\
& -\sqrt{\beta \kappa_\mathrm{ex}}\chi_\mathrm{c} \sqrt{\Gamma_\mathrm{m}}\left(j g \widetilde{\chi}_\mathrm{m}  \right) \hat{\eta}_\mathrm{m}.\\
\end{align*}
In the frequency domain, the power spectral density of the noise at the output (in units of W/Hz) 
is $S=\hbar \omega \langle\hat{b}_\mathrm{out}^\dagger \hat{b}_\mathrm{out}\rangle $,
\begin{align*}
S & = \frac{ 4 \hbar \omega \beta \kappa_\mathrm{ex}  (\Gamma_\mathrm{m}^2+ 4\delta^2)\kappa n_\mathrm{c} }{\left|4 g^2+\left(\kappa+2j (\delta+\widetilde{\Delta})\right)\left(\Gamma_\mathrm{m}+2j \delta\right)\right|^2}\\
& +\frac{ 16 \hbar \omega \beta \kappa_\mathrm{ex}  g^2 \Gamma_\mathrm{m}  n_\mathrm{m}^\mathrm{T} }{\left|4 g^2+\left(\kappa+2j (\delta+\widetilde{\Delta})\right)\left(\Gamma_\mathrm{m}+2j \delta\right)\right|^2}.
\end{align*} 

The first term simply represents the thermal noise of a cavity with occupancy $n_\mathrm{c}$ whose spectral weight is distributed over the `dressed' cavity mode.  The `dressed' cavity mode includes the effect of optomechanically induced transparency \cite{S_Agarwal2010,S_Weis2010,S_Teufel2010} and reduces to a single Lorentzian lineshape in the limit of weak coupling ($g \ll \sqrt{\kappa \Gamma_\mathrm{m}}$).  The second term is the thermal noise of the mechanical mode with its modified mechanical susceptibility.  Unlike previous derivations \cite{S_Rocheleau2010}, we have not assumed the weak-coupling regime.  Thus, as this equation is valid in both the weak- and strong-coupling regimes, it gives a unified description of the thermal noise spectrum even in the presence of normal-mode splitting.    Finally, the total noise at the output of the measurement including the vacuum noise of the photon field and the added noise of the measurement is
\begin{equation}
\frac{S}{\hbar \omega}= \frac{1}{2}+n_\mathrm{add}'+\frac{ 4 \beta \kappa_\mathrm{ex} \left[\kappa n_\mathrm{c} (\Gamma_\mathrm{m}^2+ 4\delta^2)+4 \Gamma_\mathrm{m} n_\mathrm{m}^\mathrm{T} g^2 \right]}{\left|4 g^2+\left(\kappa+2j (\delta+\widetilde{\Delta})\right)\left(\Gamma_\mathrm{m}+2j \delta\right)\right|^2},
\end{equation} 
where $n_\mathrm{add}'$ is the total added noise of the measurement in units of equivalent number of photons.  For an ideal measurement (\emph{i.e.} for a quantum-limited measurement of both quadratures of the light field), $n_\mathrm{add}'=1/2$. 

Before the onset of normal-mode splitting, one can directly relate the measured microwave power spectrum $S$ to the displacement spectral density $S_x$.  Assuming $\widetilde{\Delta}=0$, $n_\mathrm{c} \ll n_\mathrm{m}$ and $g,\delta \ll \kappa$,
\begin{align}
\frac{S}{\hbar \omega} & = \frac{1}{2}+n_\mathrm{add}'+ 4 \beta \frac{\kappa_\mathrm{ex}}{\kappa}\Gamma\frac{\Gamma_\mathrm{m} n_\mathrm{m}^\mathrm{T}}{\left(\Gamma_\mathrm{m}+\Gamma\right)^2+4\delta^2}\\
& = \frac{1}{2}+n_\mathrm{add}'+\frac{2\beta G^2 n_\mathrm{d}}{\kappa}\frac{\kappa_\mathrm{ex}}{\kappa}S_x, 
\end{align} 
where $\Gamma=4g^2/\kappa$ is the optomechanical damping rate.  

\section{Microwave measurement and calibration}

The detailed circuit diagram for our measurements is shown in Fig.~S2. In order to calibrate the value of $g_0=Gx_\mathrm{zp}$ for this device, we applied a microwave drive optimally red-detuned ($\widetilde{\Delta}=0$)  and measured the thermal noise spectrum of the mechanical oscillator as a function of cryostat temperature.  Here we restricted $n_\mathrm{d} \approx 3$ in order to ensure that radiation pressure effects are negligible.  With the value of $g_0$ now determined, we increase the drive amplitude and measure the thermal noise spectrum at each drive power.  The noise spectra are recorded and averaged with commercial FFT spectrum analyser. Each spectrum is typically an average of 500 traces with a measurement time of 0.5~s per trace.  The cavity response is then measured with a weak probe tone with a vector network analyser to determine precise cavity parameters at each microwave drive power, including the precise detuning and $\kappa$.  For larger microwave drive powers where the cavity spectrum exhibits optomechanically induced transparency effects \cite{S_Agarwal2010,S_Weis2010,S_Teufel2010}, this spectrum also serves as a direct measure of $g$.  Finally, using additional calibration tones, each noise spectrum is calibrated in units of absolute microwave noise quanta and fit with Eq.~5 to determine the occupancy of both the cavity and mechanical modes.  
  
For our measurements, we infer that our entire measurement chain has an effective added noise of $n_\mathrm{add}'=2.1$.  This value is consistent with the independently measured value for the added noise of the JPA ($n_\mathrm{add}=0.8$) and the $2.5$~dB of loss between the output of the cavity and the JPA \cite{S_Castellanos-Beltran2008,S_Teufel2009}. 

\begin{figure}[!bh]
\includegraphics[width=\columnwidth]{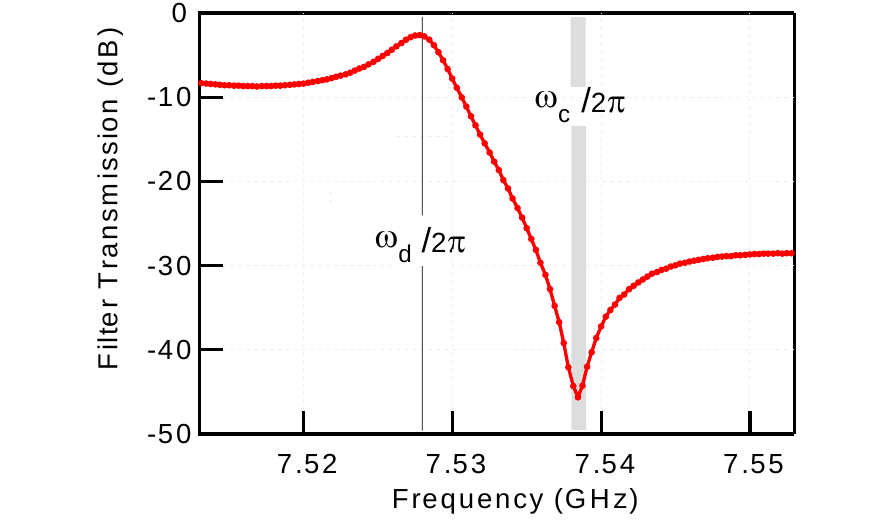}
\caption{\label{supfig3} \textbf{Measured transmission of filter cavity.}  A tunable resonant cavity was implemented at room temperature in order to suppress noise $\sim 10$~MHz above the drive frequency.  As shown here, this cavity reduces the noise at the cavity frequency by more than 40~dB, ensuring that the phase or amplitude noise of the generator is not responsible for the finite occupancy of the cavity at large drive power. }
\end{figure} 

In order to ensure that the amplitude or phase noise of the signal generator was not responsible for the finite occupancy of the cavity at high drive power, we designed and built a custom filter cavity \cite{S_Rocheleau2010}.  As shown in Fig.~S3, when the filter cavity is tuned to precisely the frequencies of our circuit, it provides an addition 40~dB of noise suppression at the cavity resonance frequency.  The phase and amplitude noise of our signal generator alone are specified by the manufacturer to be less than -150~dBc at Fourier frequencies $10$~MHz away from the drive.  With the addition of filter cavity, we lower this noise to well below the shot-noise level of our microwave drive.  Furthermore, even without the filter cavity, we could not resolve an appreciable difference in the cavity occupation.  Thus, while we do not know the precise mechanism for this occupancy, we conclusively determine the generator noise is not the cause. 

\section{Inferring cavity parameter and number of drive photons}

The measured microwave cavity parameters may be inferred from the transmitted power spectrum.  The power at the output of the cavity $P_\mathrm{o}$ is related to the input power $P_\mathrm{i}$ by \cite{S_Teufel2010}
\begin{equation}
P_\mathrm{o}=P_\mathrm{i}\left(  \frac{\kappa_\mathrm{0}^2+4\Delta^2}{\kappa^2+4\Delta^2} \right),
\end{equation} 
where $\Delta=\omega_\mathrm{d}-\omega_\mathrm{c}$ is the difference between the frequency of the drive $\omega_\mathrm{d}$ and the cavity resonance frequency $\omega_\mathrm{c}$.  $\kappa$ is the total intensity decay rate of the cavity (full width at half maximum) with $\kappa=\kappa_\mathrm{0}+\kappa_\mathrm{ex}$.  $\kappa_\mathrm{0}$  is the coupling rate to the dissipative environment, and $\kappa_\mathrm{ex}$ is the coupling rate to the transmission line used to excite and monitor the cavity. 

The number of photons in the cavity due to a coherent input drive at detuning $\Delta$ may be calculated from the stored energy $E$ in the cavity.
\begin{equation}
n_\mathrm{d}=\frac{E}{\hbar \omega_\mathrm{d}}=\frac{2 P_{\mathrm{i}}}{\hbar \omega_{\mathrm{d}}}\frac{\kappa_{\mathrm{ex}}}{\kappa^2+4\Delta^2}
\end{equation} 
For our circuit, $\kappa_\mathrm{ex}=2 \pi \times 133$~kHz. Thus, when the drive is optimally detuned such that $\Delta=-\Omega_\mathrm{m}$, the input power required to excite the cavity with one photon is $ P_\mathrm{i} \approx 2\hbar \omega_{\mathrm{d}}\Omega_\mathrm{m}^2/\kappa_\mathrm{ex} \approx 50$~ fW. 

\section{Fundamental limits of sideband cooling}

Equation 2 in the main text gives an expression for the final occupancy of a mechanical mode, assuming that the microwave drive is optimally detuned ($\Delta=-\Omega_\mathrm{m}$).  This expression is only the lowest order approximation in the small quantities $g/\Omega_\mathrm{m}$ and $\kappa/\Omega_\mathrm{m}$.  Up to second order, the final occupancy is  \cite{S_Dobrindt2008}
\begin{align*} 
n_\mathrm{m} &= n_\mathrm{m}^\mathrm{T}\left( \frac{\Gamma_\mathrm{m}}{\kappa}\frac{4g^2+\kappa^2}{4g^2+\kappa \Gamma_\mathrm{m}}\right)\left[1+ \frac{g^2}{\Omega_\mathrm{m}^2}\frac{4g^2+\kappa \Gamma_\mathrm{m}}{4g^2+\kappa^2}   \right] \\
&+ n_\mathrm{c}\left(\frac{4g^2}{4g^2+\kappa \Gamma_\mathrm{m}}\right)\left[1+ \frac{8g^2+\kappa^2}{8\Omega_\mathrm{m}^2}\frac{4g^2+\kappa\Gamma_\mathrm{m}}{4g^2}   \right] \\
&+\frac{8g^2+\kappa^2}{16\Omega_\mathrm{m}^2}.
\end{align*} 
The last term represents the fundamental limit for sideband cooling and demonstrates the importance of the resolved-sideband regime.  For our system, $\Omega_\mathrm{m} \gg \kappa, g$; and hence this last term only contributes negligibly to the final occupancy of the mechanical mode ($< 10^{-4}$ quanta). 

\subsection{Measurement imprecision and backaction} 

Throughout the main text and this supplementary information, we use the ``single-sided" convention for all spectral densities in which for any quantity $A$, the mean-square fluctuations are $\left<A^2\right>=\int_{0}^{\infty}S_A(\omega)\frac{d\omega}{2\pi}$.  This yields the familiar classical result that an oscillator coupled to a thermal bath of temperature $T$ will experience a random force characterized by the force spectral density $S_{F}=4k_B T m \Gamma_{\mathrm{m}}$.  More generally, 
\begin{equation}
S_F=4\hbar \Omega_\mathrm{m}\left(n_\mathrm{m}^\mathrm{T}+\frac{1}{2}\right)m\Gamma_\mathrm{m} \,,
\end{equation} 
where $n_\mathrm{m}^\mathrm{T}$ is the Bose-Einstein occupancy factor given by $n_\mathrm{m}^\mathrm{T}=[\exp (\hbar \Omega_\mathrm{m}/k_B T)-1]^{-1}$.  

Independent of any convention for defining the spectral density, the visibility of a thermal mechanical peak of given mechanical occupancy above the noise floor of the measurement represents a direct measure of the overall efficiency of the detection.  As shown in Fig.~S4, ratio of the peak height to the white-noise background allows us to quantify the imprecision of the measurement in units of mechanical quanta \cite{S_Teufel2009}, $n_\mathrm{imp}\equiv S_x^\mathrm{imp} m \Omega_\mathrm{m} \Gamma_\mathrm{m}'/(4 \hbar)$.  Inspection of Eq.~S6 implies 
\begin{equation}
n_\mathrm{imp} = \frac{1}{4\beta}\frac{\kappa}{\kappa_\mathrm{ex}}\frac{4 g^2+\kappa\Gamma_\mathrm{m}}{4g^2}\left(\frac{1}{2}+n_\mathrm{add}'\right)\\
\end{equation}

\begin{figure}[!bh]
\includegraphics[width=\columnwidth]{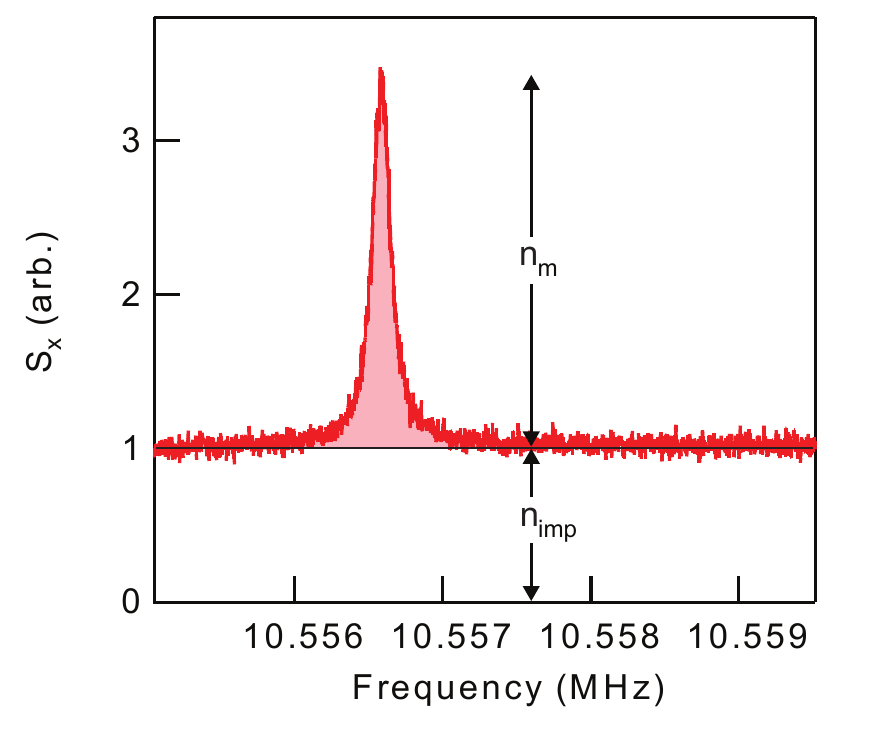}
\caption{\label{supfig4} \textbf{Measurement imprecision in units of mechanical quanta.} }
\end{figure} 

Once the drive is strong enough that $g \gg \sqrt{\kappa\Gamma_\mathrm{m}}$), $n_\mathrm{imp}$ no longer decreases with increasing drive.  It is precisely because we are measuring with a detuned drive that also damps the mechanical motion, that $n_\mathrm{imp}$ asymptotically approaches a constant value \cite{S_Clerk2010,S_Schliesser2009}.  For an ideal measurement ($\beta=1, \kappa=\kappa_\mathrm{ex}$, and $n_\mathrm{add}'=1/2$), $n_\mathrm{imp}\rightarrow 1/4$.  Implicit in obtaining this optimal value for $n_\mathrm{add}'$ and hence $n_\mathrm{imp}$ is that all the photons exiting the cavity are measured.   Any losses between the cavity and the detector can be modeled as a beam-splitter that only transmits a fraction $\eta$ of the photons to the detector and adds a fraction $(1-\eta)$ of vacuum noise.  So the effective added noise $n_\mathrm{add}'$ accounting for these losses becomes
\begin{equation}
n_\mathrm{add}'=\frac{n_\mathrm{add}}{\eta}+\left(\frac{1-\eta}{\eta}\right)\frac{1}{2},   
\end{equation}
Thus, shot-noise limited detection of the photons ($n_\mathrm{add}=1/2$) is a necessary, but not sufficient, condition for reaching the best possible level of precision. 

Quantum mechanics also requires that a continuous displacement measurement must necessarily impart a force back on the measured object.  For an optimally detuned drive ($\widetilde{\Delta}=0$) in the resolved-sideband regime, this backaction force spectral density $S_F^\mathrm{ba}$ approaches a constant value as a function of increasing drive strength and asymptotically approaches $S_F^\mathrm{ba}=2\hbar \Omega_\mathrm{m}m\Gamma_\mathrm{m}'$.  Again, expressing the spectral density in units of mechanical quanta gives $n_\mathrm{ba}\equiv S_F^\mathrm{ba}/(4\hbar \Omega_\mathrm{m}m\Gamma_\mathrm{m}')\rightarrow 1/2$.

Fundamentally, the Heisenberg limit does not restrict the imprecision $S_x^\mathrm{imp}$ or the backaction $S_F^{ba}$ alone, but rather it requires their product has a minimum value \cite{S_Braginsky1992,S_Clerk2010}
\begin{equation}
 \sqrt{S_x^\mathrm{imp} S_F^{ba}}=4 \hbar \sqrt{n_\mathrm{imp}n_\mathrm{ba}} \ge \hbar.
\end{equation}
An ideal cavity optomechanical system can achieve this lower limit for a continuous measurement with a drive applied at the cavity resonance frequency.  When considering the case where the drive is instead applied detuned below the cavity resonance ($\widetilde{\Delta}=0$), this product never reaches this lower limit \cite{S_Clerk2010,S_Schliesser2009} and is at minimum $\sqrt{S_x^\mathrm{imp} S_F^{ba}}=\hbar\sqrt{2}$.  

To estimate these quantities for our measurements, we can infer the total force spectral density experienced by our oscillator as $S_F^\mathrm{total}=4\hbar \Omega_\mathrm{m}m\Gamma_\mathrm{m}'(n_\mathrm{m}+1/2)$.  As this total necessarily includes the backaction, we may make the most conservative assumption that it was solely due to backaction that our oscillator remained at finite occupancy.  Hence, $n_\mathrm{ba} \le n_\mathrm{m} +1/2$.  
The low thermal occupancies attained in this work allow us to place an upper bound on how large the backaction could possibly be, and hence quantify our measurement in terms of approach to the Heisenberg limit.  Thus, $\sqrt{S_x^\mathrm{imp} S_F^{ba}}=4 \hbar \sqrt{n_\mathrm{imp} n_\mathrm{ba}} \le 4 \hbar\sqrt{n_\mathrm{imp}(n_\mathrm{m}+1/2)}$.   At  $n_\mathrm{d}=3 \times 10^4$,  we simultaneously achieve $n_\mathrm{m}=0.36$  and $n_\mathrm{imp}=1.9$ ($S_F^\mathrm{total}=1.6 \times 10^{-34}$~N$^2$/Hz and $S_x^\mathrm{imp}=1.7 \times 10^{-33}$~m$^2$/Hz) yielding an upper limit on the measured product of backaction and imprecision of $5.1~\hbar$.  As stated above, the best possible backaction-imprecision product is $\hbar \sqrt{2}$ when using red-detuned excitation; thus our measurement is only a factor of $3.6$ above this limit.  It may also be noted that this factor would have been $1.8$ except that our chosen geometry losses half of the signal back to the input ($\beta=1/2$).  In future experiments, using a single-port geometry ($\beta=1$) will improve this inefficiency.

\end{document}